\documentclass[preprint, superscriptaddress]{revtex4-1}
\usepackage{graphicx}
\usepackage{dcolumn}
\usepackage{amsmath}
\usepackage{amsfonts}
\usepackage{amssymb}
\usepackage{mathrsfs}
\usepackage{physics}
\usepackage{mathtools}
\usepackage{stmaryrd}
\usepackage{tabularx}
\usepackage{comment}
\usepackage[table]{xcolor}
\definecolor{lightgray}{RGB}{230,230,230}
\usepackage{caption}
\definecolor{model1color}{RGB}{201, 220, 255}
\definecolor{model2color}{RGB}{255, 206, 201}
\definecolor{model3color}{RGB}{211, 246, 233}
\definecolor{model4color}{RGB}{255, 229, 201}
\usepackage{orcidlink}
\usepackage{booktabs}
\usepackage{algorithm}
\usepackage{algpseudocode}
\usepackage{hyperref}
\hypersetup{
	colorlinks=true,
	linkcolor=blue,
	filecolor=blue,      
	urlcolor=blue,
	citecolor=blue,
	pdfstartview={XYZ null null 1.07}
}
\usepackage[T1]{fontenc}
\usepackage[latin1]{inputenc}
\usepackage{booktabs}
\usepackage[font=small,labelfont=bf,tableposition=top]{caption}
\usepackage{rotating}
\usepackage{array,multirow}
\usepackage{float}
\usepackage{adjustbox}
\usepackage{makecell} 

\begin{document}
\begin{center}
\hrule height 4.1pt
\vspace{0.5cm}
{\Large\textbf{Financial Fraud Detection: A Comparative Study of Quantum Machine Learning Models}}
\vspace{0.5cm}
\hrule height 1.2pt
\end{center}

\author{Nouhaila Innan\orcidlink{0000-0002-1014-3457}}\email[]{nouhailainnan@gmail.com}
\affiliation{\footnotesize Quantum Physics and Magnetism Team, LPMC, Faculty of Sciences Ben M'sick, Hassan II University of Casablanca, Morocco}
\affiliation{\footnotesize Quantum Formalism Fellow, Zaiku Group Ltd, Liverpool, United Kingdom}

\author{Muhammad Al-Zafar Khan\orcidlink{0000-0002-1147-7782}}\email[]{muhammadalzafark@gmail.com}
\affiliation{\footnotesize Quantum Formalism Fellow, Zaiku Group Ltd, Liverpool, United Kingdom}
\affiliation{\footnotesize Robotics, Autonomous Intelligence, and Learning Laboratory (RAIL), School of Computer Science and Applied Mathematics, University of the Witwatersrand, 1 Jan Smuts Ave, Braamfontein, Johannesburg 2000, Gauteng, South Africa}

\author{Mohamed Bennai}
\email[]{mohamed.bennai@univh2c.ma}
\affiliation{\footnotesize Quantum Physics and Magnetism Team, LPMC, Faculty of Sciences Ben M'sick, Hassan II University of Casablanca, Morocco}

\begin{abstract}
In this research, a comparative study of four Quantum Machine Learning (QML) models was conducted for fraud detection in finance. We proved that the Quantum Support Vector Classifier model achieved the highest performance, with F1 scores of $0.98$ for fraud and non-fraud classes. 
Other models like the Variational Quantum Classifier, Estimator Quantum Neural Network (QNN), and Sampler QNN demonstrate promising results, propelling the potential of QML classification for financial applications. While they exhibit certain limitations, the insights attained pave the way for future enhancements and optimisation strategies.
However, challenges exist, including the need for more efficient quantum algorithms and larger and more complex datasets. The article provides solutions to overcome current limitations and contributes new insights to the field of Quantum Machine Learning in fraud detection, with important implications for its future development.   
\end{abstract}

\maketitle

\section{Introduction}
\textit{Fraud} is the act of deceiving and misleading a person, or group of people, with the intention of obtaining some kind of gain (oftentimes financial). It involves the provisioning of misrepresented information or data to the victim, which seems ``too good to be true'', or the request of the victim's private data. Frequently, the targets of these attacks are elderly folk or those individuals whom are not technologically inclined. Fraudsters play on the emotions of their victims by usually creating a need for urgency around performing a certain task, like the victim disclosing his/her confidential information like identity/social security numbers, pin codes, One-Time Pins (OTPs), or other information that can render the victim susceptible. Over the years, fraud schemes have become even more sophisticated, and with the advent of Generative Artificial Intelligence (GenAI) becoming more ubiquitous, more suave and ultra-modern schemes such as the employment of various phishing scams and Natural Language Processing (NLP) to use voices of the victim's family members or friends are used in order to gain their trust, and credence. 

Broadly speaking, fraud can be categorised into the following flavours:
\begin{enumerate}
\item[I.1.1.] \textbf{Purloinment of Identity:} Also known as ``identity theft'', This occurs when the perpetrator steals the personal information from the victim and ``assumes their identity'' in the sense of using their details with nefarious intent: Using the victim's personal identification number, applying for any licenses, using the victim's debit/credit card details for purchasing goods or paying for services.
\item[I.1.2.] \textbf{Insurance Claims Fraud:} This occurs when the perpetrator intentionally files fallacious insurance claims or overinflates the value of losses that occurred. 
\item[I.1.3.] \textbf{Financial Fraud:} This falls under the broader category of white collar crimes and constitutes:
    \begin{enumerate}
    \item[I.1.3.1.] \textbf{Accounting Fraud:} Also known as ``crooking the books''. This involves the deliberate manipulation and misrepresentation of figures in financial statements to mislead investors and interested parties regarding the company's financial health.
    \item[I.1.3.2.] \textbf{Ponzi and Pyramidal Schemes:} These constitute schemes whereby victims outlay some capital with the promise of receiving enormously high returns in short periods of time. In these schemes, funds are taken from the late investor ``Tom'' and given to the earlier investors ``Dick'' and ``Harry''. At the end of these schemes, the late investors are not paid out the promised return, or any return whatsoever, and the so called ``expert investment manager'' disappears. 
    \item[I.1.3.3.] \textbf{Embezzlement:} This type of fraud occurs when an entrusted party in a company holds fiduciary responsibilities and abuses their power by stealing or misappropriating funds, or assets, to suit their own objectives. 
    \item[I.1.3.4.] \textbf{Insider Trading:} This occurs when a party has access to non-public, privileged information about the company and they hedge against the company's stock price rising or plummeting. This ties into corporate espionage, where spies are deployed into companies to steal trade secrets and report to them parties of interest, who use this information to take advantage of the company.
    \end{enumerate}
\item[I.1.4.] \textbf{Wire Fraud:} Using electronic media such as emails, phone calls, text messages, or personalised social media messages to hoodwink victims. Typically, scammers will act under false pretences to impersonate an agent at a bank or institution, ask the victim to transfer funds from their accounts or disclose sensitive data. In addition, these scammers play on the victims personal troubles like romance (the famous ``Nigerian Prince scam''), or the victims financial woes like lottery prize scams, or inheritance scams, the victims philanthropic nature with charity scams, the victim's need to secure employment with job offer scams, or tech support scams.
\item[I.1.5.] \textbf{Credit Fraud:} This involves the unauthorised usage for purchasing goods, paying for services, and using the victim's debit or credit cards. Typically, this would involve the scammer getting a hold of the victim's 16-digit card number, then phishing for the card's expiry date and the 3-digit Card Verification Value (CVV).
\item[I.1.6.] \textbf{Internet Fraud:} This is the collective term for online scams and phishing attacks whereby the scammer uses emails, pop-up messages, websites, and social media to get the victim to make a payment or disclose their confidential information. 
\end{enumerate}

The focus of this paper is concentrated on credit fraud. According to a 2022 study by UK Finance, fraud resulted in losses of \textsterling 1.2 bil. (sterling), and $80\%$ of app fraud originates from online solicitations. In a 2023 study published by the news agency CNBC, it is estimated that in 2022, fraud cost consumers in the US \textdollar 8.8 bil. Such high consumer costs directly correlate to economic downturns for countries and, thus, translate to worldwide economic collapse. Thus, an accurate and quick fraud detection system is needed to tame this type of fraud. 

The idea of fraud detection using (Classical) Machine Learning (CML) models is not novel and oftentimes forms a standard textbook exercise/capstone project in this regard, and many big corporates across the financial, telecommunications, and consulting industries have fraud detection models deployed into production. For example, several of these CML models that utilise: Multivariate Logistic Regression (see \hyperlink{4}{Alenzi \& Aljehane, 2020}), Support Vector Machines (SVMs) -- see \hyperlink{9}{Kumar \textit{et al}, 2022}; \hyperlink{10}{Gyamfi \& Abdulai, 2018}, Random Forest Classifiers (see \hyperlink{33}{Liu \textit{et al}, 2015}; \hyperlink{34}{Xuan \textit{et al}, 2018a}; \hyperlink{32}{Xuan \textit{et al}, 2018b}), Gradient Boosting Machines (see \hyperlink{35}{Taha \& Malebary, 2020}), comparative studies across methods (see \hyperlink{5}{Kumar \textit{et al}, 2020}; \hyperlink{7}{Han \textit{et al}, 2020}; \hyperlink{6}{Afriyie \textit{et al}, 2023}), or combining models in ensembles (see \hyperlink{8}{Nandi \textit{et al}, 2022}) show high fidelity, robustness, and ease of implementation. 

Additionally, researchers have also applied various Deep Learning (DL) approaches: Autoencoders and Restricted Boltzmann Machines (RBMs) -- see \hyperlink{36}{Pumsirirat \& Yan, 2018}, Graph Neural Networks (GNNs) -- see \hyperlink{37}{Ma \textit{et al}, 2021}. The only time-consuming aspect of the model lifecycle is data cleaning and feature engineering. 

\textit{Quantum Machine Learning} (QML) is a newly developing field in which researchers began to express interest back in the early 2000s by combining the then emerging field of Quantum Computing (QC), an idea accredited to \hyperlink{27}{Feynman, 1982}, and CML. The goal is to leverage properties of the fundamental units of QC, qubits, and QML algorithms to obtain a computational advantage over analogous classical approaches.

However, the crystallisation and commercialisation of these ideas began to flourish in the early 2010s, and one of the most pioneering books and papers is credited to \hyperlink{28}{Wittek, 2014} and \hyperlink{29}{Biamonte \textit{et al}, 2017} respectively, who set the stage for a formalised research track -- Of course, if one looks deep enough, one may find many earlier papers, but it is beyond the scope of mentioning research works of chronological order, rather those with the highest impact. Potentially, QML can radically transform the paradigm and approach to CML by facilitating the discovery of novel algorithms that are more efficient than their classical counterparts. Since this is a rapidly developing field and we are in the Noisy Intermediate-Scale Quantum (NISQ) era of QC -- see \hyperlink{30}{Preskill, 2018}, there is no single approach. We discuss these approaches in \hyperlink{approaches_to_qml}{Tab. I.} below.

\begin{table}[H] 
\hypertarget{approaches_to_qml}{}
\caption{Approaches to Quantum Machine Learning}
\centering
\begin{tabular}{p{6.55cm}p{10.45cm}}
\hline
\hline
\textbf{Approach} &\textbf{Description} \\
\hline
Quantum Approach to CML &This entails the development of novel Quantum algorithms to solve computationally-expensive CML tasks. For example, the Quantum Support Vector Classifier has been shown to train on large datasets faster than the classical Support Vector Machine.   \\
Quantum-supplemented Approach to CML &This involves using Quantum principles to enhance existing CML algorithms. For example, the Quantum Neural Network (QNN) offers several advantages over the classical Neural Network (CNN). \\
Composite Classical-Quantum Machine Learning &This approach offers a hybrid procedure that combines elements from classical computing and QC to solve CML tasks. For example, a Quantum Computer may be used to preprocess the data, and a CML algorithm may be used to optimise the model's weights, biases, and additional parameters. \\
Applications of QML to Other Domains Besides CML &This approach involves developing and modifying existing QML algorithms for applications in areas beyond CML. As an example, QML is used extensively in the field of Computational Chemistry. One such use case is by \hyperlink{31}{Innan \textit{et al}, 2023} in which a Variational Quantum Eignesolver (VQE) was modified to perform electronic structure calculations, and a novel algorithm was presented. \\
\hline 
\end{tabular}
\end{table}

It is important to note that while QML has immense potential, it is still in the early stages of its development. Breakthroughs in hardware design, computing power, Quantum cloud technologies, and new approaches to QC will result in the more widespread adoption of QML to solve daily tasks, much like how CML is a tool that all major companies are trying to integrate and embed into their organisational processes.

The question arises: ``If these CML models are so successful and doing such a fantastic job in flagging fraudulent use cases, what is the need for QML fraud detection models?'' We advocate for adopting a Quantum approach because we believe it provides the following advantages over the classical approaches in the post-NISQ era:
\begin{enumerate}
\item[I.2.1.] \textbf{Analysis of Real-time Data:} Quantum Computers provide the opportunity to analyse vast swathes of real-time data in a methodical and structured manner with the potential to be exponentially faster. This is particularly important in fraud detection applications, where real-time detection is mandatory to mitigate the risk of large losses. 
\item[I.2.2.] \textbf{Decrease in the Amount of Inessential Data:} Fraud detection involves the analyses of large swathes of data, and although fraud accounts for such large losses, it is rare to detect while it is in progress (usually detected after it occurs), and the training data has to be specifically fabricated from real-time data; thus, a lot of redundancies occur. Since Quantum Computers offer the opportunity to analyse data in a reduced amount of time, the amount of redundant data is thereby minimised. 
\item[I.2.3.] \textbf{Scalability through Parallelisation:} QML offers the opportunity to work with larger datasets because of its ability to parallelise algorithms in a streamlined manner as compared to CML.
\item[I.2.4.] \textbf{Reduction in Algorithm Computational Complexity:} By utilising the Quantum Mechanical properties of Superposition and Entanglement, QML algorithms are less expensive than CML algorithms.
\end{enumerate}

In this paper, we apply the Quantum Support Vector Classifier, the Variational Quantum Classifier, the Estimator Quantum Neural Network, and the Sampler Quantum Neural Network to the BankSim dataset. This paper is divided into the following sections:

In \hyperlink{LitRev}{Sec. II.}, we provide a comprehensive pr\'{e}cis of the relevant literature papers pertaining to anomaly detection and fraud prediction.

In \hyperlink{Theory}{Sec. III.}, we provide an overview of the theoretical constructs of the methods used. Namely, the data encoding and the QML methods respectively.

In \hyperlink{ResDis}{Sec. IV.}, we discuss the dataset used and present the results of applying the QML models. Thereafter, we discuss the results by alluding to the various model heuristic metrics.

In \hyperlink{Concl}{Sec. V.}, we provide closing remarks on the findings of this paper. 

\section{Literature Review} \hypertarget{LitRev}{}
Since the launch of IBM's \hyperlink{41}{\texttt{Qiskit}} package and Xanadu's \hyperlink{40}{\texttt{PennyLane}}, it has become more common to apply QML methods for fraud detection. However, we note that this is a fairly new application for QML, with many papers not being very old. In this regard, we note the following literature pieces: 

Although strictly not a paper that applies the methods to fraud detection in financial data, anomaly detection forms an integral component of fraud detection. Thus, it is noteworthy to mention the work of \hyperlink{13}{Liu \& Rebentrost, 2018}, who discuss the potential applications of anomaly detection to Quantum data and propose a Quantum anomaly detection algorithm based on autoencoders. This is particularly useful when real-world data is converted to Quantum states via some feature map embedding. The research highlights the usage of the Quantum methods (Quantum Principal Component Analysis, Quantum Density Estimation, Quantum Support Vector Machines, and Quantum $k$-Nearest Neighbours) and compares them to their classical counterparts. Lastly, it gives advantages for the superiority of the Quantum methods over classical methods for anomaly detection, such as faster processing time of the data and enhanced accuracy.

\hyperlink{17}{Liang \textit{et al}, 2019} propose two Quantum anomaly detection algorithms that find applications in fraud detection. The basis for these algorithms comprises density estimation and multivariate Gaussian distributions. The goal is to find the probability density function for the training data. The advantage of this approach over classical approaches is that these algorithms scale logarithmically with respect to the number of datapoints in the training data and the dimensionality of the Quantum states. Thus, making the algorithm superior in efficiency for handling high-dimensional data. In addition, the authors propose a method for calculating the determinant of any Hermitian operator, which is particularly useful for anomalous data with a higher-dimensional normal distribution. The advantages of these algorithms are demonstrated experimentally by illustrating comparable accuracy and precision in a shorter time. 

\hyperlink{18}{Kottmann, \textit{et al}, 2021} introduced the unsupervised QML algorithm known as \textit{Variational Quantum Anomaly Detection} (VQAD) that takes simulation data and extracts the phase diagram, \textit{a priori}, without knowledge of the system. Importantly, the authors have demonstrated that the algorithm works in realistic scenarios for both real-noise simulations and on a real Quantum computer. Further, it was shown to improve the anomaly detection scheme by employing measurement error mitigation and adopting the circuits according to the physical device. Although more oriented towards Physics, the findings of this paper have potentially important implications for fraud detection. 

\hyperlink{11}{Kyriienko \& Magnusson, 2022} develop a Quantum protocol for anomaly detection and apply their technique for detecting credit card fraud. By establishing classical benchmarks, a comparative study is done against different types of Quantum kernels (products of data-dependent rotations with variational circuits, and evolution circuits, the spin-glass Hamiltonian's or the Heisenberg Hamiltonian) is established, and it is shown that Quantum fraud detection is superior to classical methods. Specifically, for supervised fraud detection, Quantum kernels offer higher expressivity and generalisability by outperforming RBF kernels, $K(\mathbf{x},\mathbf{x}')=\exp\left(-\frac{||\mathbf{x}-\mathbf{x}'||^{2}_{2}}{2\sigma^{2}}\right)$, for the free parameter $\sigma$, by over 10\% on the average precision heuristic. For unsupervised fraud detection, Quantum kernels offer a $15\%$ increase in average precision and grow as the system size grows. Lastly, the authors discuss future improvements in near- and mid-term Quantum hardware. 

\hyperlink{12}{Grossi \textit{et al}, 2022} use the \texttt{Qiskit} software stack (IBM Safer Payments and IBM Quantum Computers) to present an end-to-end application of Quantum Support Vector Machines for classification in financial services and a comparative study of the state-of-the-art QML methods collated against the classical methods. The paper shows that the hybrid method outperforms the classical method with respect to accuracy and the false positive rate (FPR) measures. Feature selection plays a pivotal role in optimising the fraud detection system. The paper proposes a Quantum Feature Importance Selection Algorithm (QFISA) that selects the most important features from a dataset to reduce the dimensionality of the dataset for running the experiment on a real Quantum device. Lastly, the drawbacks and limitations of the Factorial Analysis of Mixed Data (FAMD) method are highlighted (overlap between components, and not showing any discrimination power between the reduced variables), and it is shown how the method proposed is superior in this regard. 

\hyperlink{14}{Wang \textit{et al}, 2022} propose a framework using QML for analysing online transaction data that is time series-based, highly imbalanced, and high-dimensional in order to detect fraudulent records. Using an enhanced-Support Vector Machine with Quantum annealing solvers, they benchmark this method against CML models. This research highlights the challenges encountered when dealing with real-time transactional data and how a Quantum approach potentially provides a better approach and can be more broadly applied to other critical business applications. While providing a roadmap for further research, the authors caution that several factors must be accounted for when implementing a fraud detection model on such data; namely:
\begin{itemize}
\item \textbf{Accuracy:} How close to the actual values does one want the predicted values to be?
\item \textbf{Speed:} How urgently do you need the model to detect anomalies?
\item \textbf{Cost of Computing:} Whether one, or the company that one works for, has the financial resources to purchase hardware, and access extra qubits, to perform such calculations. 
\end{itemize}
\hyperlink{16}{Guo \textit{et al}, 2022} propose an Anomaly Detection based on the Density Estimation (ADDE) algorithm, which hinges on the estimation of the amplitude of a Quantum state, and they show that it has an exponential speed-up in the number of training datapoints and dimensions over classical algorithms. Further, the authors show how the proposed algorithm can be used for anomaly detection based on Kernel Principal Component Analysis (KPCA). Lastly, it is indicated that the findings in this paper are not limited to fraud detection but can also be applied to other domains, namely: Military surveillance, intrusion detection, and healthcare. 

Other references are contained therein in the aforementioned literature pieces. One may expect that there exists a plethora of application-based papers of QML papers for fraud detection, unexpectedly, there are not so many.  

\section{Theory} \hypertarget{Theory}{}
We present the theory of the data encoding methods used in the paper, namely: \texttt{ZZFeatureMap}, \texttt{PauliFeatureMap}, \texttt{ZFeatureMap}, and QML models: QSVC, VQC, EQNN, SQNN, used below. This is because the theory is not widely known, it helps to establish the context, justifies the choice of methods used, guides the analyses and interpretation, and enhances the overall credibility of this research. 

\subsection{Data Encoding Methods}
\label{sec:dataencoding}
\subsubsection{\texttt{ZZFeatureMap}}
\begin{figure}[htbp]
    \centering
    \hypertarget{fig1}{}
    \includegraphics[width=1\textwidth]{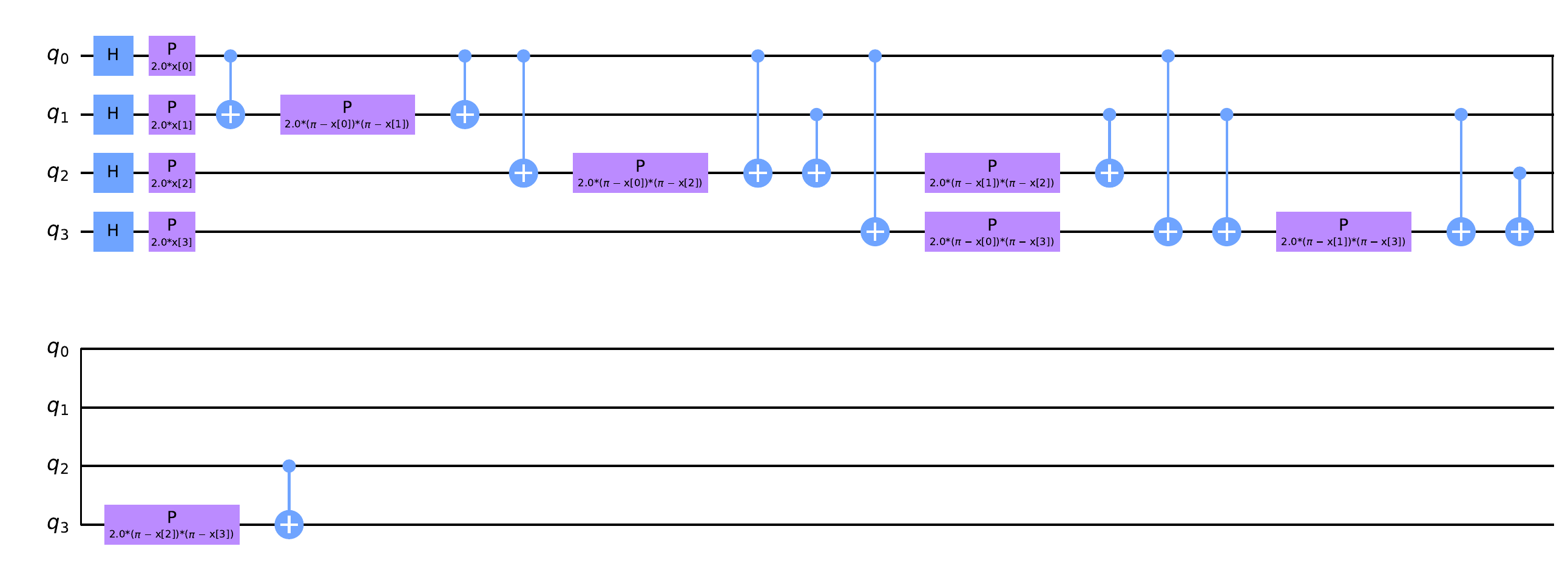} 
    \caption{Circuit diagram for the \texttt{ZZFeatureMap}.}
    \label{fig:zz}
\end{figure}

The \texttt{ZZFeatureMap class} is a Quantum circuit representing a second-order Pauli-$Z$ evolution. It takes as input a feature dimension, which is the number of qubits in the circuit, and the number of repetitions, which specifies how many times the rotation and entanglement blocks are repeated. The circuit is constructed by applying Hadamard gates to all qubits, followed by rotation and entanglement blocks as shown in \hyperlink{fig1}{Fig~\ref{fig:zz}}.

The rotation blocks apply single-qubit rotations based on the classical data, parameterised by angles determined by a classical non-linear function $\phi$, which by default is $\phi(x) = x$ for a single feature and $ \phi(x, y) = (\pi - x)(\pi - y)$ for two features, and in our case with four features: $\phi(x, y, z, w) = (\pi - x)(\pi - y)(\pi - z)(\pi - w)$. The entanglement blocks entangle the qubits based on the specified entanglement structure using controlled-$X$ (CNOT) gates.
\vspace{-1cm}
\subsubsection{\texttt{PauliFeatureMap}}
\begin{figure}[htbp]
    \centering
    \hypertarget{fig2}{}
    \includegraphics[width=1\textwidth]{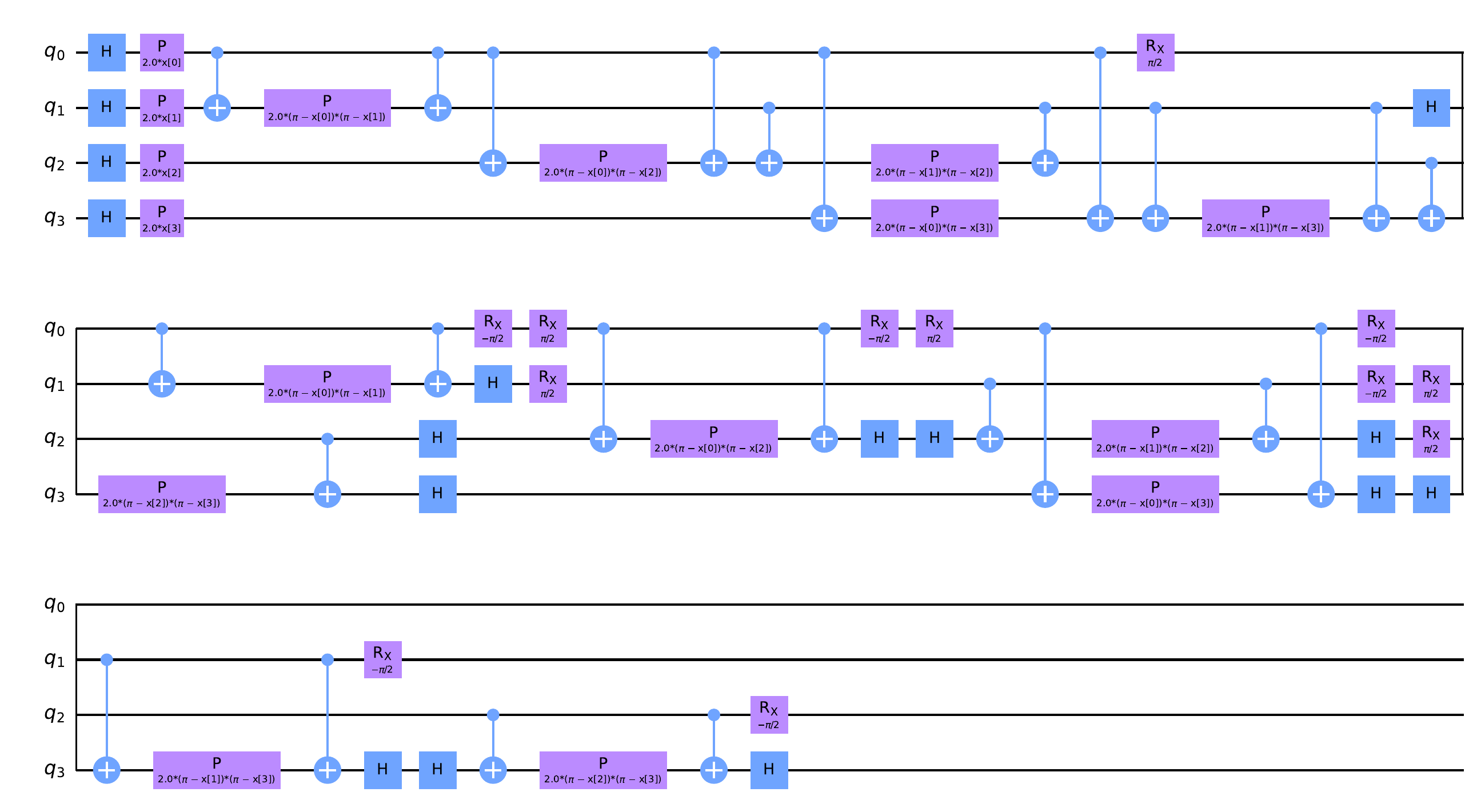} 
    \caption{Circuit diagram for the \texttt{PauliFeatureMap}.}
    \label{fig:fz}
\end{figure}

The \texttt{PauliFeatureMap} class represents a Quantum circuit that enables a Pauli expansion of a given data set. The Pauli expansion is a method for representing the data set as a product of Pauli operators, where each Pauli operator corresponds to a distinct feature within the data. The expression for the Pauli operator combination is given as:
\[U_{\varphi(\mathbf{x})} = \exp\left(\imath\sum_{S \in \mathcal{I}} \phi_S(\mathbf{x})\prod_{i\in S} P_i\right),\]
where \(\mathcal{I}\) is the set of qubit indices describing the connections in the feature map, and \(\phi_S(\mathbf{x})\) is the data mapping function. The data mapping function \(\phi_S(\mathbf{x})\) maps classical input data \(\mathbf{x}\) into the Quantum circuit, enhancing the circuit's representation capabilities. It is defined as follows:
\[\phi_S(\mathbf{x}) = \begin{cases}
x_i & \text{if } S = \{i\}, \\
\prod_{j \in S}\left(\pi - x_j\right) & \text{if } |S| > 1.
\end{cases}\]
The \texttt{PauliFeatureMap} circuit, as shown in \hyperlink{fig2}{Fig~\ref{fig:fz}}, is constructed by initially applying Hadamard gates to all qubits. Subsequently, a series of rotation gates are applied to the qubits, with the rotation angle for each qubit determined by the data function, $\phi$. Finally, entangling gates are applied to the qubits, similar to the procedure used in the previous feature map. The \texttt{PauliFeatureMap} circuit can be repeated multiple times to enhance the accuracy of the approximation, similar to other feature maps.
\vspace{-1cm}
\subsubsection{\texttt{ZFeatureMap}}

The \texttt{ZFeatureMap} class represents a first-order Pauli $Z$-evolution circuit. As a sub-class of \texttt{PauliFeatureMap}, it operates with fixed Pauli strings ``$Z$'', resulting in the absence of entangling gates in its first-order expansion. This unique characteristic makes the \texttt{ZFeatureMap} particularly well-suited for specific applications where a shallow Quantum circuit without entanglement is desired.

\begin{figure}[htbp]
    \centering
    \hypertarget{fig3}{}
    
    \includegraphics[width=0.3\textwidth]{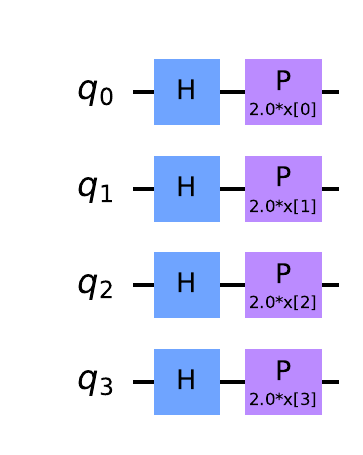} 
    \caption{Circuit diagram for the \texttt{ZFeatureMap}.}
    \label{fig:z}
\end{figure}

Similar to the \texttt{ZZFeatureMap}, the \texttt{ZFeatureMap} is tailored for a designated number of qubits, known as the \textit{feature dimension}, and the user can specify the number of repetitions to replicate the rotation blocks. The circuit is constructed by applying Hadamard gates to all qubits, followed by rotation blocks as shown in \hyperlink{fig3}{Fig~\ref{fig:z}}. The rotation blocks are structured following the same principles employed in the \texttt{ZZFeatureMap}.

The \texttt{ZFeatureMap} class also offers essential attributes for inspecting the circuit, including the feature dimension, the number of repetitions, and the entanglement strategy. In the case of the \texttt{ZFeatureMap}, the entanglement strategy is null since no entangling gates are present.

The \texttt{ZFeatureMap} class complements the \texttt{ZZFeatureMap} by providing an alternative Quantum feature map that aligns with specific use cases where entangling gates are to be avoided. Its customisable nature, and absence of entanglement, allow for efficient Quantum data encoding and processing.

\subsection{Quantum Support Vector Classifiers}

\begin{figure}[H]
    \centering
    \hypertarget{fig4}{}
    \includegraphics[width=1\linewidth]{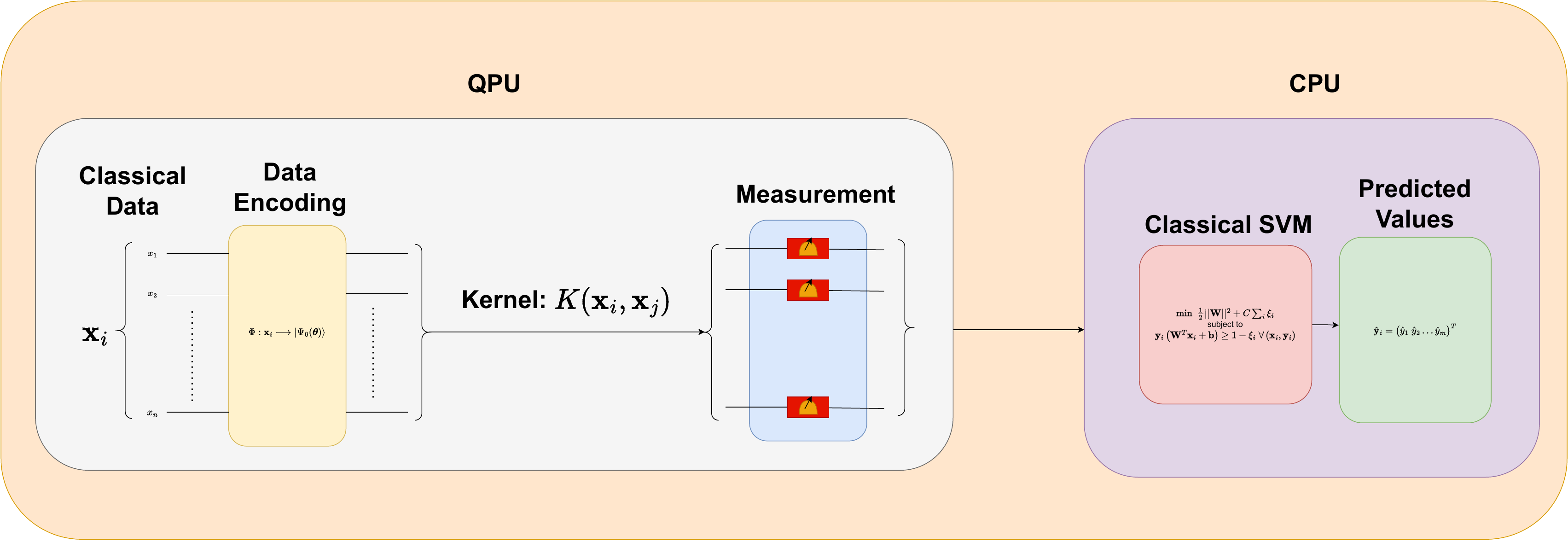}
    \caption{Architecture of the Quantum Support Vector Classifier.}
    \label{fig:qsvc}
\end{figure}

The \textit{Quantum Support Vector Classifier} (QSVC) is the Quantum Mechanical analogue of the classical Support Vector Machine (SVM), as depicted in \hyperlink{fig4}{Fig~\ref{fig:qsvc}}. The SVM model aims to find the optimal \textit{planum separans} (separating hyperplane) that categorises the datapoints. This is achieved by \textit{maximal margin classification}: Minimising the margin (distance between classes of datapoints) while simultaneously maximising the distance between the closest datapoints from each class and the hyperplane; see the excellent texts of \hyperlink{20}{Bishop, 2006}; \hyperlink{19}{Goodfellow \textit{et al}, 2016} for a full mathematical elucidation. 

The output of a QSVC is given by
\begin{equation*}
f(\mathbf{x})=\sum_{j=1}^{n}\alpha_{j}K(\mathbf{x}, \mathbf{x}_{j})+\mathbf{b},
\end{equation*}
where $\alpha_{j}$ are the coefficients of the classifier, $\mathbf{b}$ are the bias terms, and $K$ are the kernels -- which gives a measure of similarity between the datapoints $\mathbf{x}$, and the $j^{\text{th}}$ datapoint $\mathbf{x}_{j}$. \hyperlink{26}{Schuld \& Petruccione, 2021} provide an excellent discussion of the various kernel types. We advocate that the kernel is the most important component of a QSVC and significantly affects its performance. Thus, in the style of ``hyper-parameter tuning'', one should experiment with various kernels to see which gives the best model performance.

\subsection{Variational Quantum Classifiers}
\begin{figure}[H]
    \centering
    \hypertarget{fig5}{}
    \includegraphics[width=1\linewidth]{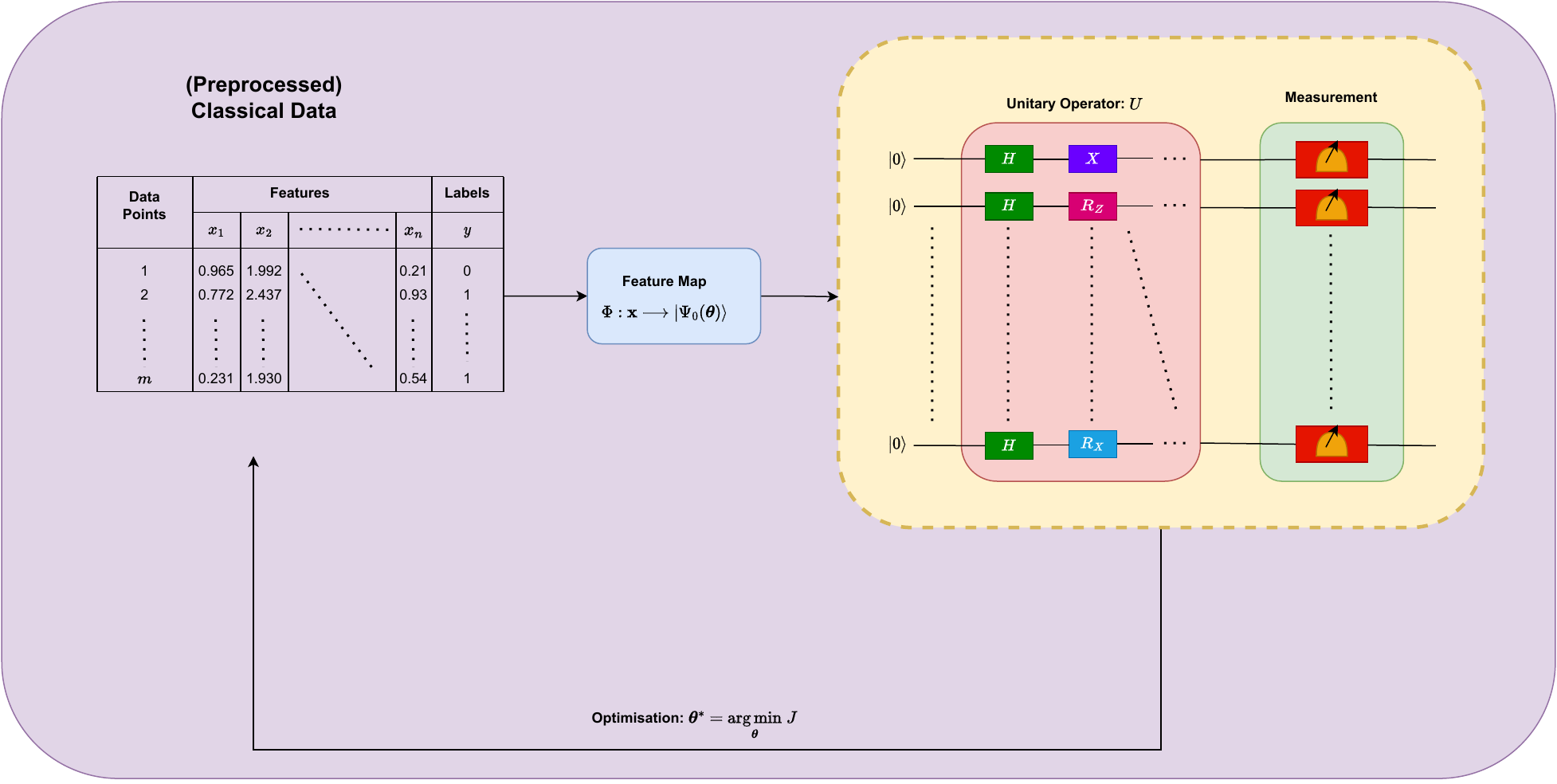}
    \caption{Architecture of the Variational Quantum Classifier.}
    \label{fig:vqc}
\end{figure}

The \textit{Variational Quantum Classifier} (VQC) is a type of Quantum circuit parameterised by learnable weights. The weights are optimised using a classical method to minimise the loss function. As indicated in \hyperlink{fig5}{Fig~\ref{fig:vqc}}, the VQC operates as follows:
\begin{enumerate}
\item[III.C.1.] \textbf{Quantum State Preparation:} Let $\boldsymbol{\theta}=\left(\theta_{1}, \theta_{2},\ldots, \theta_{n}\right)$, where $n$ is the number of registers in the circuit, be the set of learnable weights, initialised randomly for each $0\leq\theta_{i}\leq 1$. The initial state is represented as $\ket{\Psi_{0}(\boldsymbol{\theta})}$, and is oftentimes simply-prepared Quantum states such as a series $\ket{0}$ states.
\item[III.C.2.] \textbf{Application of a Unitary Transformation:} In this part of the circuit, a series of Quantum gates are applied to the initial states. \newline
Let $G_{i}\in\left\{I,X,Y,Z,H,S,T,R_{X},R_{Y},R_{Z}, \text{CNOT}, \text{SWAP},\ldots\right\}$ 
be Quantum gates, for $1\leq i\leq m$, and then we apply a series of Quantum gates on the initial state. We can be sure that no matter what combination of these Quantum gates we have, they form a unitary operator, i.e. $U=\bigotimes_{i=1}^{m}G_{i}$. Mathematically, this part of the circuit is given by $U\ket{\Psi_{0}(\boldsymbol{\theta})}\equiv \ket{\Psi(\boldsymbol{\theta})}$. 
\item[III.C.3.] \textbf{Measurement:} Measurement is performed on the result $U(\boldsymbol{\theta})$ in order to extract information from the Quantum states.
\end{enumerate}
Steps III.C.2. and III.C.3. are repeated in order to minimise the loss function, $J(\boldsymbol{\theta})$, and the process is stopped once an acceptance criterion is met. 

\subsection{Estimator Quantum Neural Networks}
\begin{figure}[H]
    \centering
    \hypertarget{fig6}{}
    \includegraphics[width=1\linewidth]{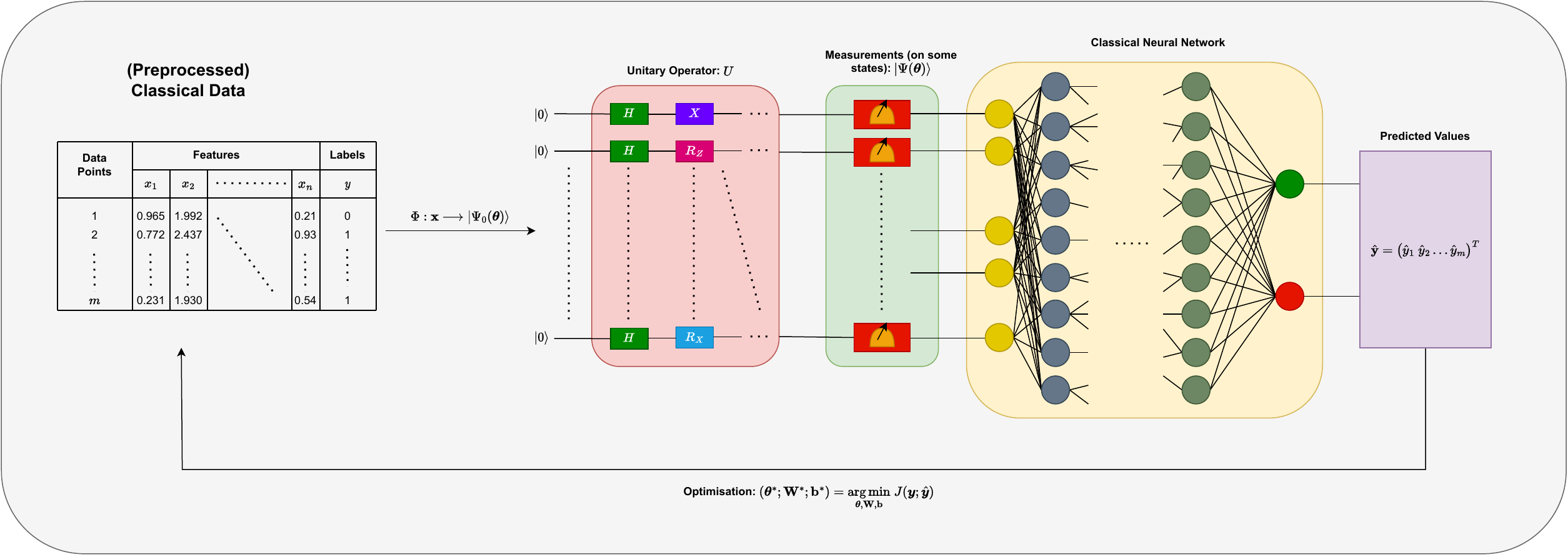}
    \caption{Architecture of the Estimator QNN.}
    \label{fig:eqnn}
\end{figure}

The \textit{Estimator Quantum Neural Network} (EQNN) is a hybrid Classical-Quantum neural Network architecture whereby the Quantum component is known as the \textit{feature map} and converts the classical data into Quantum states. As shown in \hyperlink{fig6}{Fig~\ref{fig:eqnn}}, the EQNN operates as follows:
\begin{enumerate}
\item[III.D.1.] \textbf{State Preparation via Quantum Feature Map:} \hypertarget{EQNN_state_prep}{} Given classical data $\mathbf{x}=\left(x_{1},x_{2},\ldots,x_{n}\right)$, the Quantum feature map, $\Phi:\mathbf{x}\longrightarrow\ket{\Psi_{0}(\boldsymbol{\theta})}$, encodes the classical data into parameterised Quantum states, $\ket{\Psi_{0}(\boldsymbol{\theta})}$, using the VQC. As is the case with the VQC, the states $\ket{\Psi_{0}(\boldsymbol{\theta})}$ are oftentimes just a series of $\ket{0}$ states.
\item[III.D.2.] \textbf{Performing Measurement:} Measurement is performed on (some of) the qubits in the computational ($\left\{\ket{0}, \ket{1}\right\}$) basis to obtain classical features. 
\item[III.D.3.] \textbf{Processing in a Classical Neural Network:} The classical features that are extracted here are fed to fully-connected classical neural network architecture in order to produce the predicted values, $\hat{\mathbf{y}}$. 
\item[III.D.4.]\textbf{Model Optimisation and Optimal Parameter Search:} In this step, the architecture is optimised to discover the optimal parameters $\boldsymbol{\theta}^{*}$ of the VQC, as well as the weights, $\mathbf{W}^{*}$, and biases, $\mathbf{b}^{*}$, of the classical neural network, such that the loss function is minimised; i.e. $\left(\boldsymbol{\theta}^{*};\mathbf{W}^{*}; \mathbf{b}^{*}\right)=\underset{\boldsymbol{\theta},\mathbf{W},\mathbf{b}}{\arg\min}\;J(\mathbf{y};\hat{\mathbf{y}})$. Importantly, this optimal search is carried out in parallel. 
\end{enumerate}

\subsection{Sampler Quantum Neural Networks}
\begin{figure}[H]
    \centering
    \hypertarget{fig7}{}
    \includegraphics[width=1\linewidth]{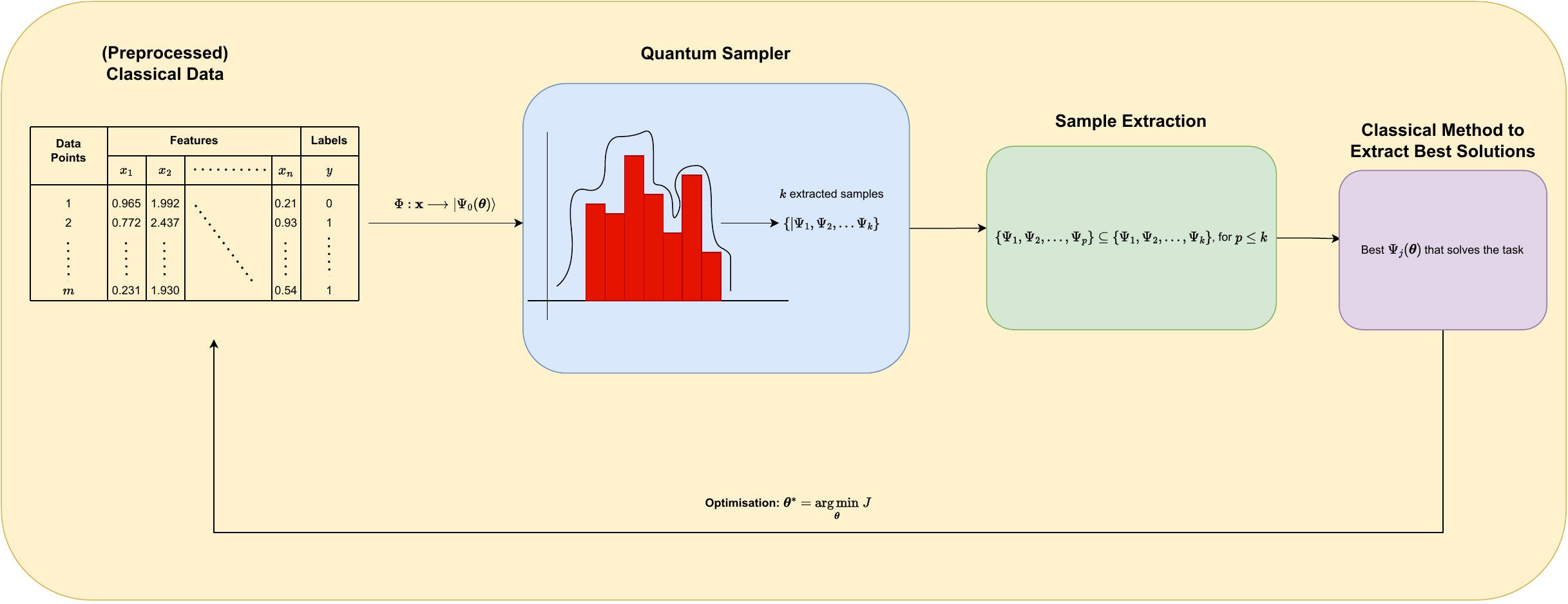}
    \caption{Architecture of the Sampler QNN.}
    \label{fig:sqnn}
\end{figure}

Analogous to the EQNN, the \textit{Sampler Quantum Neural Network} (SQNN) also contains a hybrid Classical-Quantum architecture. However, the SQNN is equipped with a \textit{Quantum Sampler}, which extracts example Quantum states from the complex probability distributions associated with the Quantum states. As illustrated  in \hyperlink{fig7}{Fig~\ref{fig:sqnn}}, the SQNN operates as follows:
\begin{enumerate}
\item[III.E.1.] \textbf{State Preparation via Quantum Feature Map:} Exactly the same as the case of the EQNN; see \hyperlink{EQNN_state_prep}{III.D.1.} 
\item[III.E.2.] \textbf{Application of the Quantum Sampler:} Oftentimes this is taken to be the Quantum Approximate Optimisation Algorithm (QAOA); see \hyperlink{25}{Farhi \textit{et al}, 2014}. The purpose is to efficiently extract example Quantum states from the complex probability distribution corresponding to problem solutions under specific variable configurations.
\item[III.E.3.] \textbf{Sample Extraction:} Samples are chosen from the examples generated by the Quantum sampler.
\item[III.E.4.] \textbf{Utilising Classical Methods to Extract the Best Solutions:} From the samples, the best solutions to the given task are chosen using some kind of classical scheme. 
\item[III.E.5.] \textbf{Optimal Parameter Search:} The optimal parameters are found using a classical optimisation method in order to minimise the cost function, i.e. $\boldsymbol{\theta}^{*}=\underset{\boldsymbol{\theta}}{\arg\min}\;J$.The values of $\boldsymbol{\theta}$ are fed back to the VQC, and the process begins once again. The process is repeated until the optimal values of the parameters are found. 
\end{enumerate}

\section{Results and Discussion} \hypertarget{ResDis}{}
\subsection{Dataset and Feature Selection}
The dataset used in this research study is derived from \hyperlink{22}{BankSim}, an agent-based simulator of bank payments based on aggregated transactional data provided by a prominent bank in Spain. The primary objective of BankSim is to generate synthetic data tailored explicitly for fraud detection research. To achieve this goal, statistical analysis and Social Network Analysis (SNA) were deployed to study the relationships between merchants and customers, developing a calibrated model -- see \hyperlink{21}{Lopez-Rojas \& Axelsson, 2014}.

The BankSim dataset encompasses $594\;643$ records obtained over $180$ steps, simulating approximately six months of temporal activity. From these records, $587\;443$ are regular payments, while $7\;200$ are classified as fraudulent transactions. It is important to note that the simulated fraud occurrences were introduced by incorporating thieves aiming to steal an average of three cards per step and performing around two fraudulent transactions per day.

The dataset comprises nine feature columns and one target column, each offering essential insights to discern underlying patterns and characteristics. The features encompassed are as follows:
\begin{itemize}
    \item \textbf{Step:} Representing the temporal aspect, this feature denotes the simulation day, effectively encompassing $180$ steps, emulating six months.
    \item \textbf{Customer:} Denoting customer identification, this feature distinguishes individual customers engaging in transactions.
\item \textbf{ZipCodeOrigin:} An indicator of each transaction's zip code of origin or source, offering the potential for geographic analysis.
    \item \textbf{Merchant:} Capturing the merchant's identification, this feature differentiates between various merchants involved in the transactions.
    \item \textbf{ZipMerchant:} This feature denotes the zip code associated with each merchant, providing further potential for geographic insights.
    \item \textbf{Age:} Representing the customer's age, this feature is categorized into discrete age groups, including ``$0$''$:(\leq 18),\quad$``$1$''$:(19-25),\quad$``$2$'' $:(26-35),\quad$``$3$''$: (36-45),$ ``$4$''$: (46-55),\quad$``$5$''$: (56-65),\quad$``$6$''$: (> 65),$ $\text{and ``U''}: \text{(Unknown)}$.
    \item \textbf{Gender:} Categorizing the gender of each customer, this feature includes values such as ``E'' (Enterprise), ``F'' (Female), ``M'' (Male), and ``U'' (Unknown).
    \item \textbf{Category:} Capturing the category of each purchase transaction, this feature imparts valuable insights into the nature and type of transactions.
    \item \textbf{Amount:} Representing the monetary value of each purchase, this feature offers critical information on transaction volumes.
    \item \textbf{Fraud:} This binary target variable classifies each transaction as fraudulent (denoted by ``1'') or benign (denoted by ``0''). This classification forms the basis for the subsequent fraud detection analysis.
\end{itemize}

Graphical analysis played a crucial role in deepening our understanding of the dataset. We generated several visualisations, including histograms, bar plots, and a heatmap, to gain valuable insights into the data distribution and uncover potential patterns.

\begin{figure}[htbp]
    \centering
    \hypertarget{fig8}{}
    \includegraphics[width=0.8\textwidth]{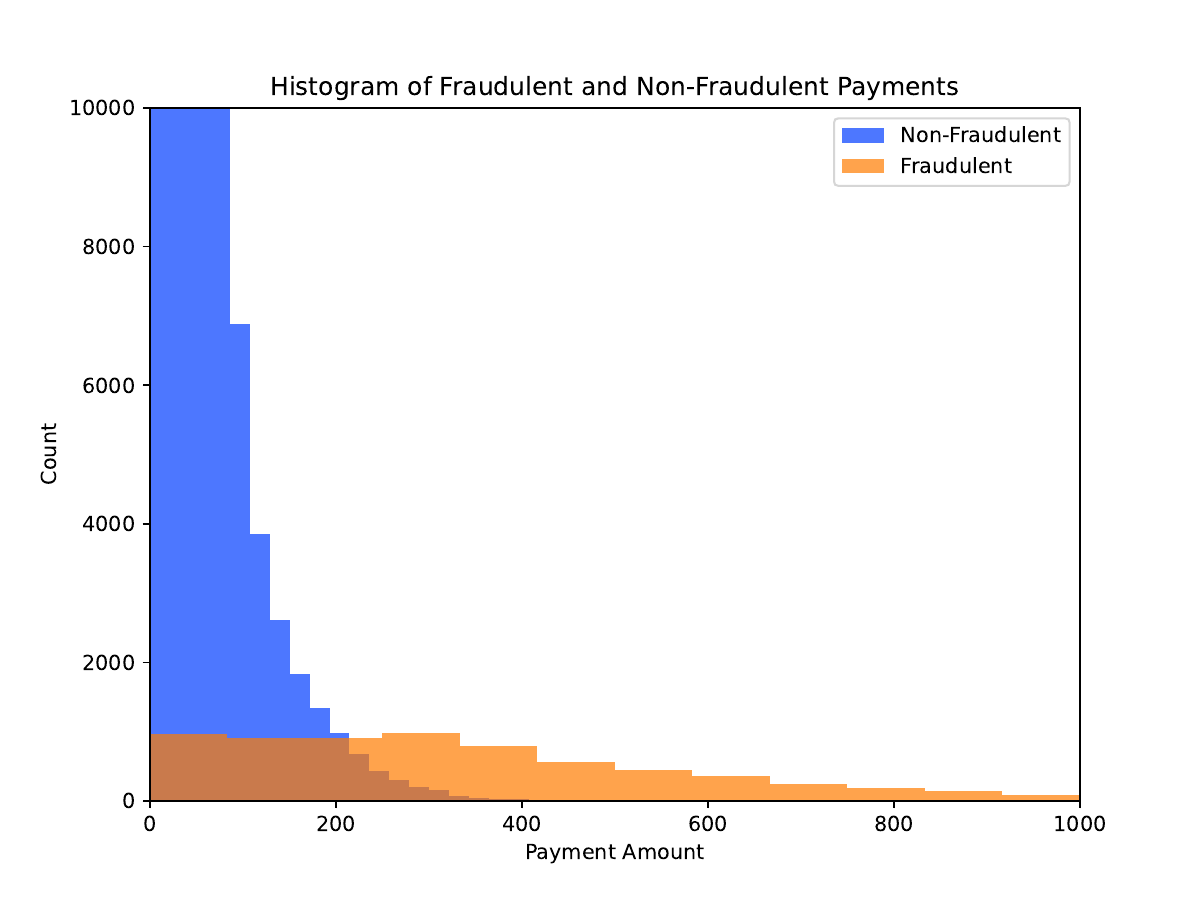} 
    \caption{Histogram of fraudulent and non-fraudulent payments.}
    \label{fig:f5}
\end{figure}
\hyperlink{fig8}{Fig.~\ref{fig:f5}} displays a histogram comparing payment amounts for fraudulent and non-fraudulent transactions. Our analysis reveals that fraudulent transactions involve higher payment amounts on average (mean = $567.23$, std = $128.47$) compared to legitimate transactions (mean = $145.68$, std = $50.32$). This insight highlights the significance of payment amount as a distinguishing factor between the two transaction categories.

\begin{figure}[htbp]
    \centering
    \hypertarget{fig9}{}
    \includegraphics[width=0.8\textwidth]{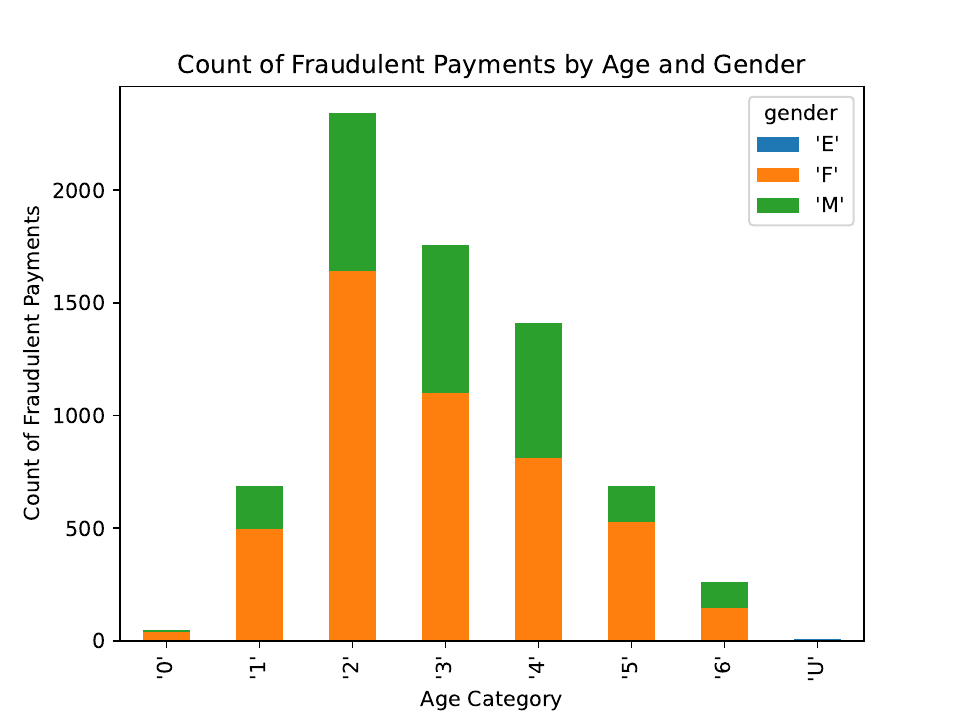} 
    \caption{Count of fraudulent payments by age and gender.}
    \label{fig:f3}
\end{figure}

\hyperlink{fig9}{Fig.~\ref{fig:f3}} presents a bar plot depicting fraudulent payments categorized by age and gender. The visualisation indicates that individuals aged $26$ to $35$ $(45\%)$ and females $(56\%)$ constitute more fraudulent transactions. In comparison, males $(34\%)$ and individuals aged $36$ to $45$ years $(32\%)$ show a lower incidence of involvement in fraudulent activities. These demographic trends offer valuable guidance for developing targeted fraud detection strategies.

\begin{figure}[htbp]
    \centering
    \hypertarget{fig10}{}
    \includegraphics[width=0.8\textwidth]{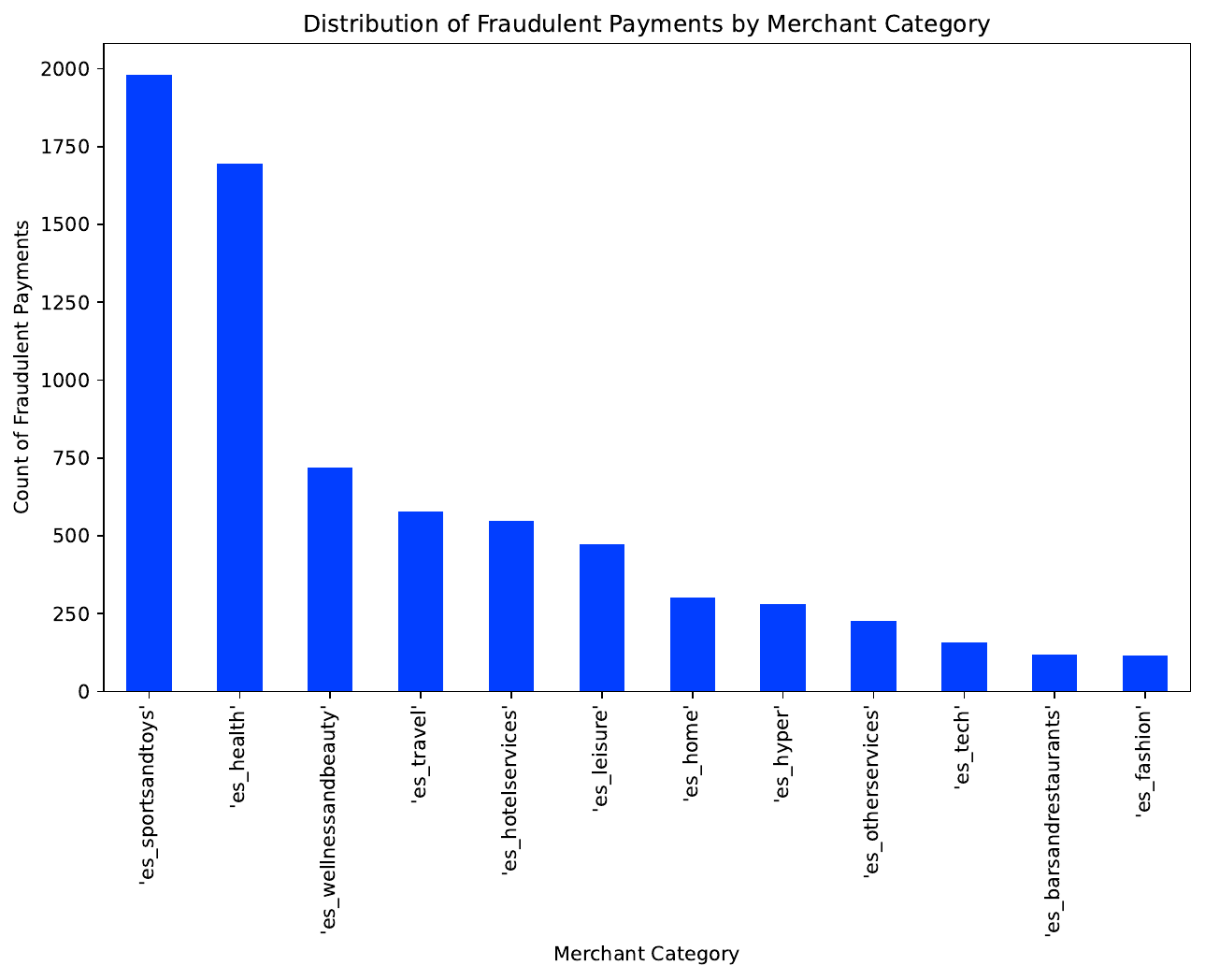} 
    \caption{Distribution of fraudulent payments by merchant category.}
    \label{fig:f4}
\end{figure}

\hyperlink{fig10}{Fig.~\ref{fig:f4}} illustrates the distribution of fraudulent payments across different merchant categories. Specific merchant categories, such as ``sports \& toys'' and ``health'', exhibit a disproportionately higher occurrence of fraudulent transactions, representing $20\%$ and $15\%$ of all fraud cases, respectively. This finding emphasizes the importance of considering merchant categories as a relevant feature in our fraud detection models.

\begin{figure}[htbp]
    \centering
    \hypertarget{fig11}{}
    \includegraphics[width=0.8\textwidth]{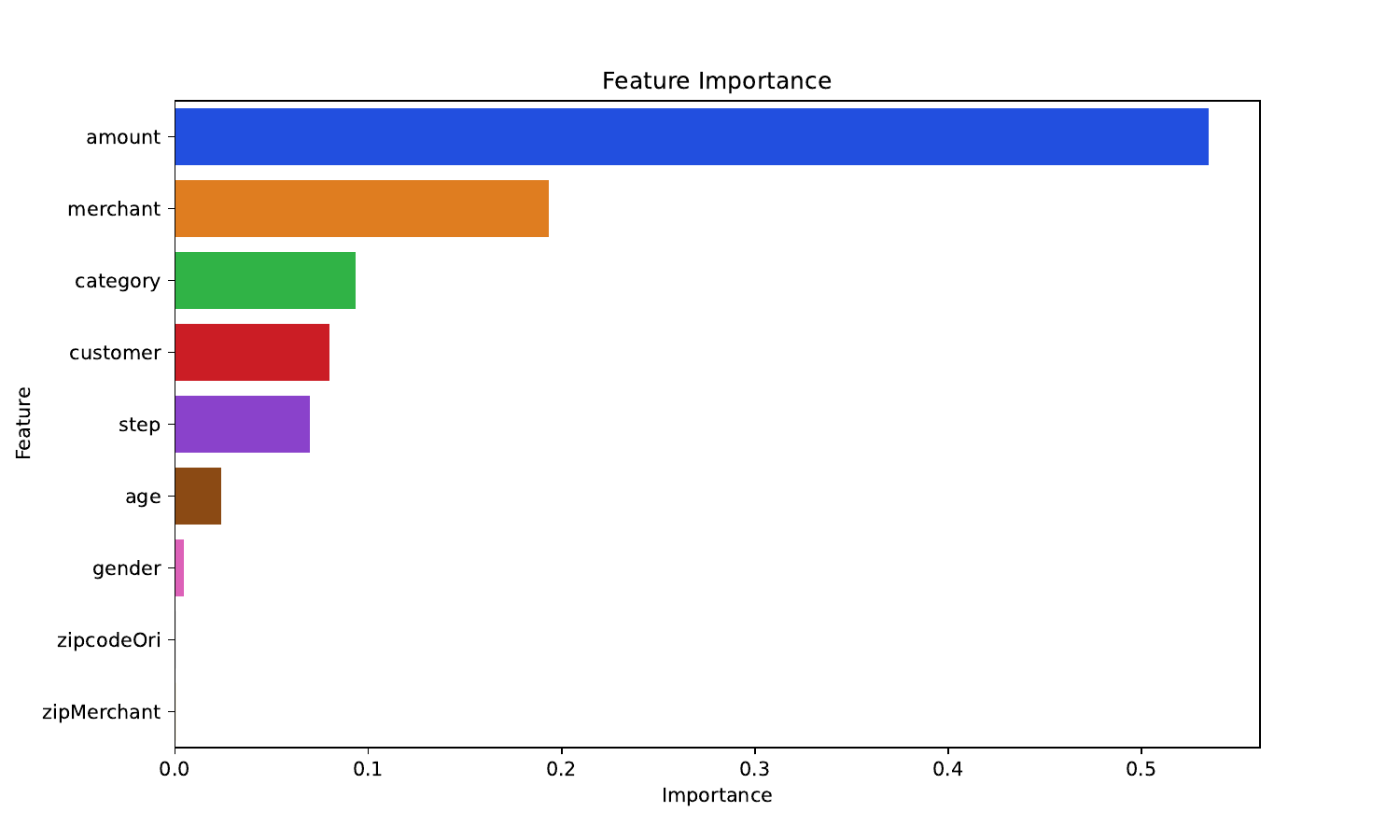} 
    \caption{Feature importance in fraud detection.}
    \label{fig:f1}
\end{figure}

To identify the most informative features that significantly contribute to our fraud detection models, we employed Principal Component Analysis (PCA) to reduce the dimensionality of the dataset while preserving the most valuable information. As shown in \hyperlink{fig11}{Fig.~\ref{fig:f1}}, the results of the PCA analysis indicated the order of importance of the features based on their corresponding principal components. Notably, the feature ``amount'' emerged as the most influential, followed by ``merchant,'' ``category,'' ``customer,'' ``step,'' ``age,'' ``gender,'' ``zipMerchant,'' and ``zipcodeOri.'' This valuable ranking guided our further feature selection process.
Subsequently, we conducted a logical analysis to investigate the relationships between these selected features and their potential impact on fraud detection. The logical analysis confirmed that the features ``age,'' ``gender,'' `category,'' and ``amount'' exhibited distinct patterns in fraudulent and non-fraudulent transactions, making them promising candidates for our fraud detection models.

\begin{figure}[htbp]
    \centering
    \hypertarget{fig12}{}
    \includegraphics[width=0.8\textwidth]{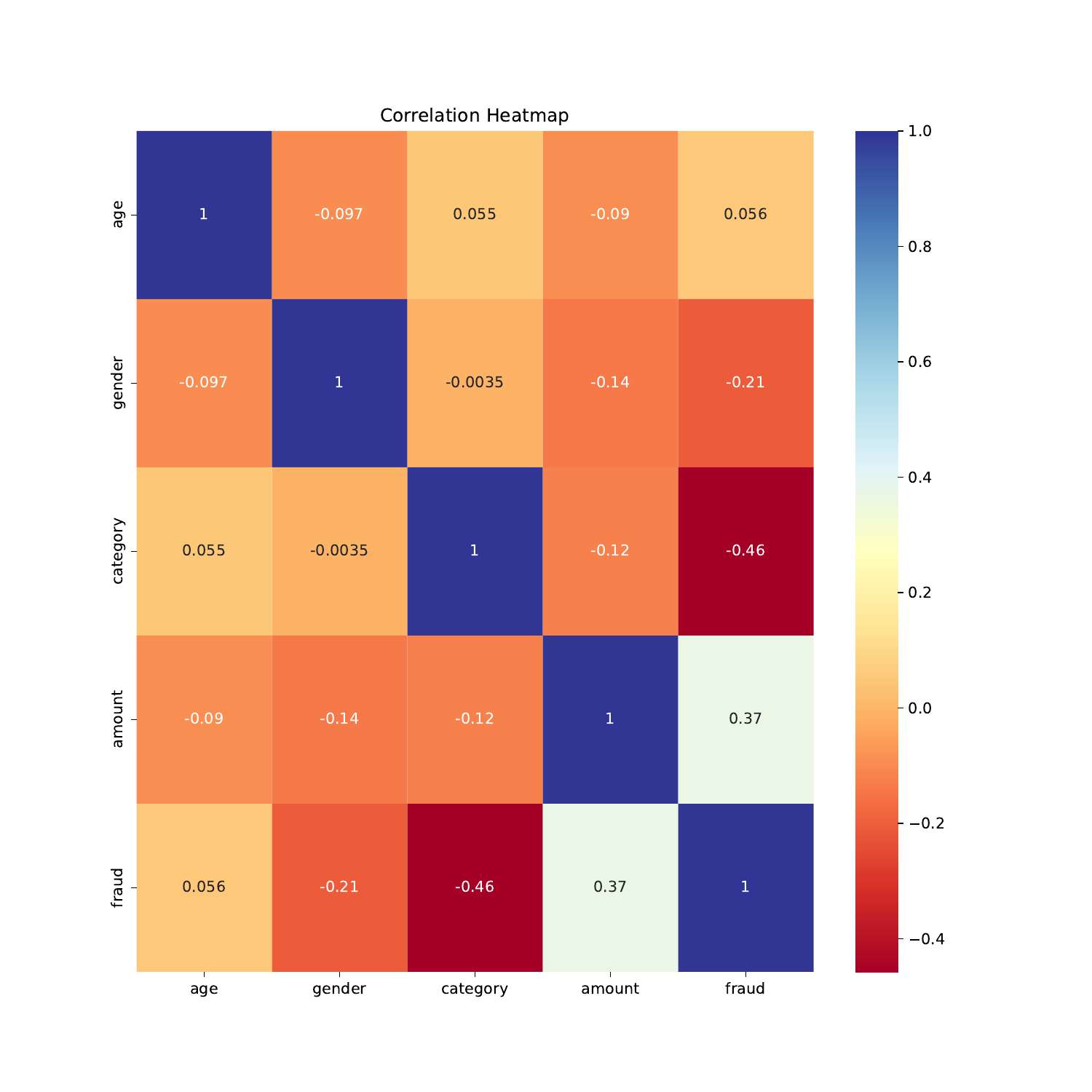} 
    \caption{Correlation heatmap of features.}
    \label{fig:f2}
\end{figure}

We further examined the correlation heatmap to gain deeper insights into the relationships among the selected features (\hyperlink{fig12}{Fig.~\ref{fig:f2}}). The heatmap matrix displayed the pairwise correlations among ``age,'' ``gender,'' ``category,'' ``amount,'' and ``fraud.''
The correlation heatmap showcased the strength and nature of the relationships. Notably, the feature ``amount'' exhibited a weak negative correlation with ``fraud,'' suggesting a potential association between higher transaction amounts and fraudulent transactions.
Based on the insights gained from the logical analysis and confirmed by the correlation heatmap, we concluded that the features ``age,'' ``gender,'' ``category,'' and ``amount'' were the most informative variables for our fraud detection models. Incorporating these features into our fraud detection framework allows us to deliver robust and efficient financial security and risk management practices, advancing the field.

\subsection{Data Analysis and Experimental Setup}
Before conducting the fraud detection analysis, a rigorous data preprocessing and cleaning process was undertaken to ensure the dataset's quality and suitability for reliable model training and evaluation. The original dataset was loaded, and specific subsets were extracted to create a balanced dataset containing $200$ records with $100$ instances of fraudulent and non-fraudulent transactions.
A data transformation step addressed inconsistencies in the ``age'' column, which contained non-numeric characters, by extracting numerical values from the age categories using regular expressions. Consequently, each age was converted to an integer for accurate representation in the subsequent analysis.

To prepare the dataset for model training, certain categorical features, such as ``category'' and ``gender,'' were transformed into numerical representations using \texttt{scikit-learn}'s LabelEncoder. This encoding process allowed the model to process these categorical variables effectively during training. Subsequently, the dataset was further prepared by removing unused features and converting the remaining features into numerical values to ensure homogeneity across the data.
The dataset was split into training and testing sets using the \texttt{train\_test\_split} function from \texttt{scikit-learn} to facilitate the model training process. The training set denoted as $X_{\text{train}}$ and $y_{\text{train}}$ contained a portion of the data used for training the model. The testing set, represented as $X_{\text{test}}$ and $y_{\text{test}}$, was kept separate and served as unseen data to evaluate the model's performance.

The feature matrix $X$ encompassed all pertinent features, excluding the ``fraud'' column, which served as the target variable. The target variable, denoted as $y$, distinguished between fraudulent transactions (encoded as $(1)$) and non-fraudulent transactions (encoded as $(0)$). This distinction was essential for the model to learn patterns and accurately classify new data. Following these preprocessing steps and dividing the dataset into training and testing sets, the data was ready for the subsequent model training and evaluation processes.

We employed our four Quantum Machine Learning models for the training process, each tailored to specific configurations. To optimize these models effectively, we harnessed the power of the Qiskit optimizer, implementing the COBYLA algorithm with a maximum iteration limit of $200$. This prudent choice of optimizer facilitated efficient convergence towards the optimal solution, ensuring the training process was effective and resource-efficient.

To provide an ideal environment for training, we utilized the Aer backend with the QasmSimulator. This choice enabled us to simulate Quantum circuits effectively, enabling seamless training of the models.
Following the training process, we meticulously evaluated the performance of each model using various key metrics. These metrics comprehensively understood each model's predictive capabilities and effectiveness.

\subsection{Results and Interpretation}
In this paragraph, we present a comprehensive evaluation of our Quantum Machine Learning models, including QSVC, VQC, EQNN, and SQNN, on our dataset using three distinct feature maps: \texttt{ZZFeatureMap}, \texttt{PauliFeatureMap}, and \texttt{ZFeatureMap}. The primary evaluation metrics were precision, recall, and F1 scores for the fraud (Class 1) and non-fraud (Class 0) cases.

The results demonstrated that the QSVC model utilizing the \texttt{ZFeatureMap} achieved the highest performance, with impressive F1 scores of $0.98$ for fraud and non-fraud classes \hyperlink{performance}{Tab. II}. Furthermore, the QSVC model accurately identified fraudulent and non-fraudulent transactions.
\begin{table}[htbp]
\hypertarget{performance}{}
\caption{Performance Comparison of Quantum Machine Learning Models}
\centering
\hspace{-1.5cm}
\begin{tabularx}{\textwidth}{l|ccc|ccc|ccc}
\toprule
\textbf{QML Model} & \multicolumn{3}{c|}{\textbf{ZZFeatureMap}} & \multicolumn{3}{c|}{\textbf{PauliFeatureMap}} & \multicolumn{3}{c}{\textbf{ZFeatureMap}} \\
\cmidrule(lr){2-4} \cmidrule(lr){5-7} \cmidrule(lr){8-10}
& \textbf{Precision} & \textbf{Recall} & \textbf{F1-score} & \textbf{Precision} & \textbf{Recall} & \textbf{F1-score} & \textbf{Precision} & \textbf{Recall} & \textbf{F1-score} \\
\midrule

\rowcolor{model1color}
\textbf{QSVC} & & & & & & & & & \\
Class 0 & 0.62 & 0.68 & 0.65 & 0.58 & 0.61 & 0.59 & 1.00 & 0.97 & 0.98 \\
Class 1 & 0.68 & 0.62 & 0.65 & 0.56 & 0.52 & 0.54 & 0.97 & 1.00 & 0.98 \\
\rowcolor{lightgray}
Accuracy & \multicolumn{3}{c|}{0.65} & \multicolumn{3}{c|}{0.56} & \multicolumn{3}{c}{0.98} \\
\rowcolor{lightgray}
Macro avg & 0.65 & 0.65 & 0.65 & 0.57 & 0.57 & 0.56 & 0.98 & 0.98 & 0.98 \\
\rowcolor{lightgray}
Weighted avg & 0.65 & 0.65 & 0.65 & 0.57 & 0.57 & 0.57 & 0.98 & 0.98 & 0.98 \\

\midrule

\rowcolor{model2color}
\textbf{VQC} & & & & & & & & & \\
Class 0 & 0.55 & 0.52 & 0.53 & 0.54 & 0.45 & 0.49 & 0.86 & 0.95 & 0.9 \\
Class 1 &  0.52 & 0.55 & 0.53 & 0.50 & 0.59 & 0.54 & 0.93 & 0.84 & 0.88 \\
\rowcolor{lightgray}
Accuracy & \multicolumn{3}{c|}{0.53} & \multicolumn{3}{c|}{0.52} & \multicolumn{3}{c}{0.90} \\
\rowcolor{lightgray}
Macro avg & 0.53 & 0.53 & 0.53 &  0.52 & 0.52 &  0.52 & 0.89 & 0.9 & 0.9 \\
\rowcolor{lightgray}
Weighted avg & 0.53 & 0.53 & 0.53 & 0.52 & 0.52 & 0.51 & 0.89 &  0.9 &  0.9\\

\midrule

\rowcolor{model3color}
\textbf{EstimatorQNN} & & & & & & & & & \\
Class 0 & 0.53 & 0.52 & 0.52 & 0.52 & 0.45 & 0.48 & 0.70 & 1.00 & 0.83 \\
Class 1 & 0.50 & 0.52 & 0.51 & 0.48 & 0.55 & 0.52 & 1.00 & 0.55 & 0.71 \\
\rowcolor{lightgray}
Accuracy & \multicolumn{3}{c|}{0.52} & \multicolumn{3}{c|}{0.50} & \multicolumn{3}{c}{0.78} \\
\rowcolor{lightgray}
Macro avg & 0.52 & 0.52 & 0.52 & 0.50 & 0.50 & 0.50 & 0.85 & 0.78 & 0.77 \\
\rowcolor{lightgray}
Weighted avg & 0.52 & 0.52 & 0.52 & 0.50 & 0.50 & 0.50 & 0.85 & 0.78 & 0.77 \\

\rowcolor{model4color}
\textbf{SamplerQNN} & & & & & & & & & \\
Class 0 &  0.57 & 0.68 & 0.62 & 0.52 & 0.45 & 0.48 & 0.58 & 0.71 & 0.64 \\
Class 1 & 0.57 & 0.45 & 0.50 & 0.48 & 0.55 & 0.52 & 0.59 & 0.45 & 0.51 \\
\rowcolor{lightgray}
Accuracy & \multicolumn{3}{c|}{0.57} & \multicolumn{3}{c|}{0.50} & \multicolumn{3}{c}{0.58} \\
\rowcolor{lightgray}
Macro avg & 0.57 & 0.56 & 0.56 & 0.50 & 0.50 & 0.50 & 0.58 & 0.58 & 0.57 \\
\rowcolor{lightgray}
Weighted avg & 0.57 & 0.57 & 0.56 & 0.50 & 0.50 & 0.50 & 0.58 & 0.58 & 0.58 \\
\bottomrule
\end{tabularx}
\label{tab:tab1}
\end{table}

Notably, the QSVC model, based on the Qiskit library's QSVC class, does not involve a loss function as in classical machine learning algorithms. Instead, it leverages the Quantum kernel to measure the similarity between Quantum states, enabling it to classify data points effectively.

\begin{figure}[htbp]
    \centering
    \hypertarget{fig13}{}
    \includegraphics[width=0.85\textwidth]{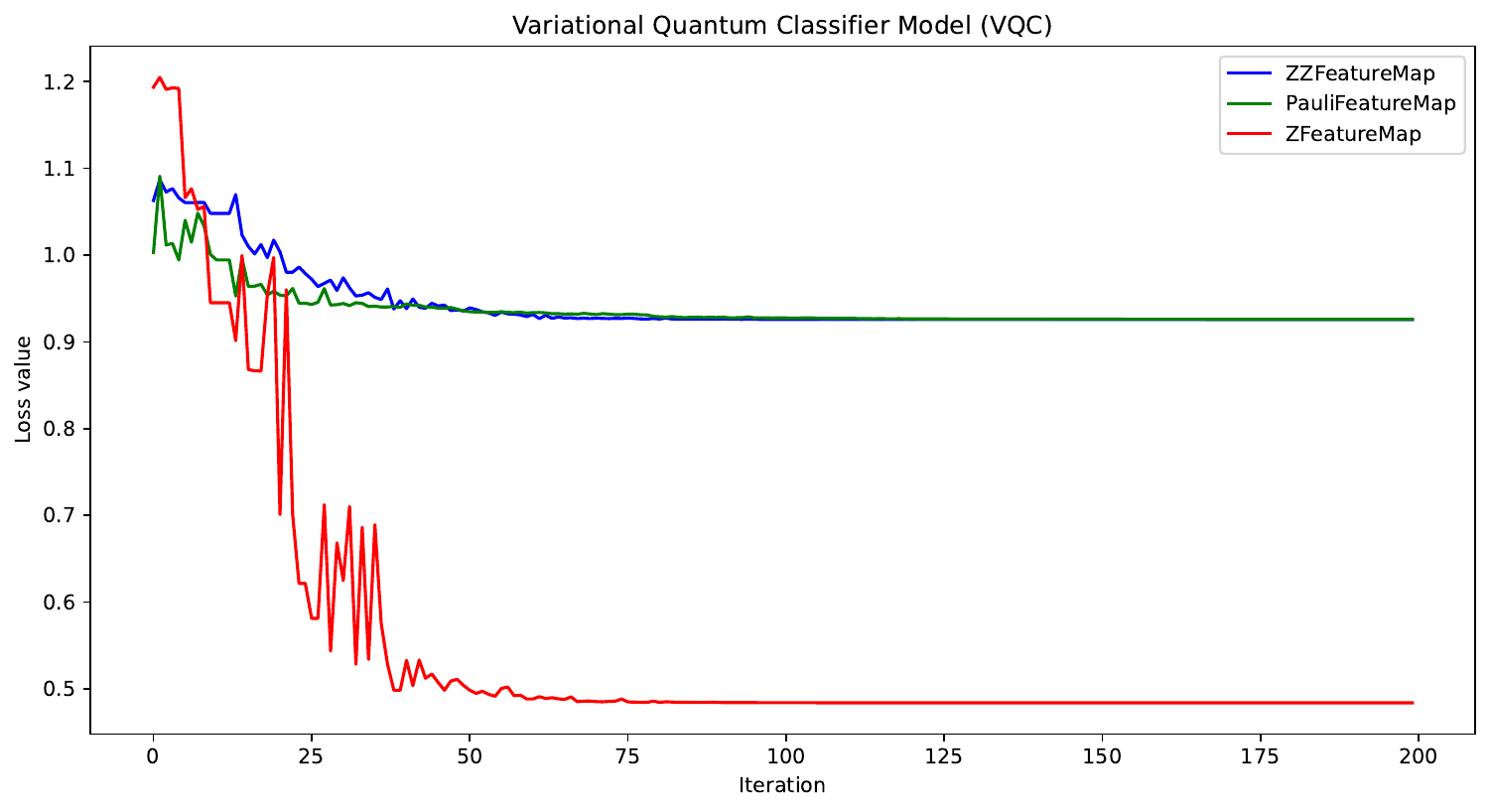} 
    \caption{Loss function of Variational Quantum Classifier model.}
    \label{fig:vqcloss}
\end{figure}

The VQC model, also employing the \texttt{ZFeatureMap}, performed well with an F1 score of $0.90$. However, in contrast to QSVC, the VQC model experienced a loss during training. In \hyperlink{fig13}{Fig.~\ref{fig:vqcloss}}, we observe that the VQC model achieved a loss of $0.5$ when using the \texttt{ZFeatureMap}, while losses of $0.95$ were observed for the \texttt{PauliFeatureMap} and \texttt{ZZFeatureMap}. The lower loss with \texttt{ZFeatureMap} indicates that this data encoding strategy leads to better convergence during the optimisation process, contributing to the higher accuracy achieved by the VQC model with this feature map.

\begin{figure}[htbp]
    \centering
    \hypertarget{fig14}{}
    \includegraphics[width=0.85\textwidth]{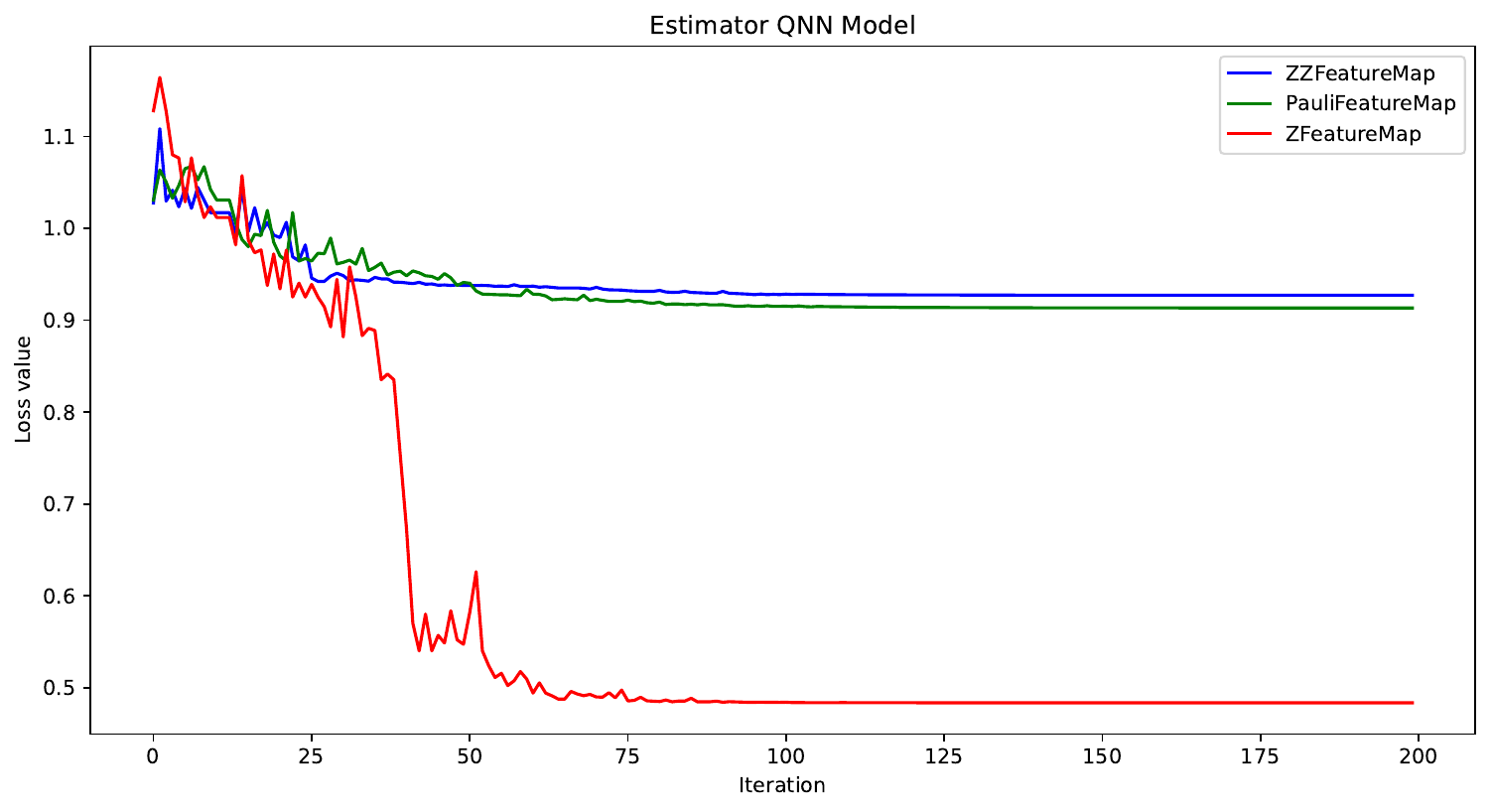} 
    \caption{Loss function of Estimator QNN model.}
    \label{fig:eqnnloss}
\end{figure}

On the other hand, the EQNN model, using the \texttt{ZFeatureMap}, showed a relatively lower F1 score of $0.78$. \hyperlink{fig14}{Fig.~\ref{fig:eqnnloss}} illustrates the corresponding loss values, with the \texttt{ZFeatureMap} achieving a loss of $0.5$, the \texttt{PauliFeatureMap} a loss of $0.96$, and the \texttt{ZZFeatureMap} a loss of $0.97$. The higher losses with the latter two feature maps suggest that the optimisation process encountered difficulties reaching an optimal solution, reducing accuracy for the EQNN model.

The limited accuracy of the EQNN model might result from the inherent limitations of the Quantum circuits used for data encoding. The simplicity of the Quantum circuit utilized by the EQNN model might not adequately capture the complex patterns present in the dataset. Exploring more expressive Quantum circuits or advanced Quantum architectures could offer potential improvements.
Similarly, the SQNN model demonstrated lower accuracy than the other models, which was expected. 

\begin{figure}[htbp]
    \centering
    \hypertarget{fig15}{}
    \includegraphics[width=0.85\textwidth]{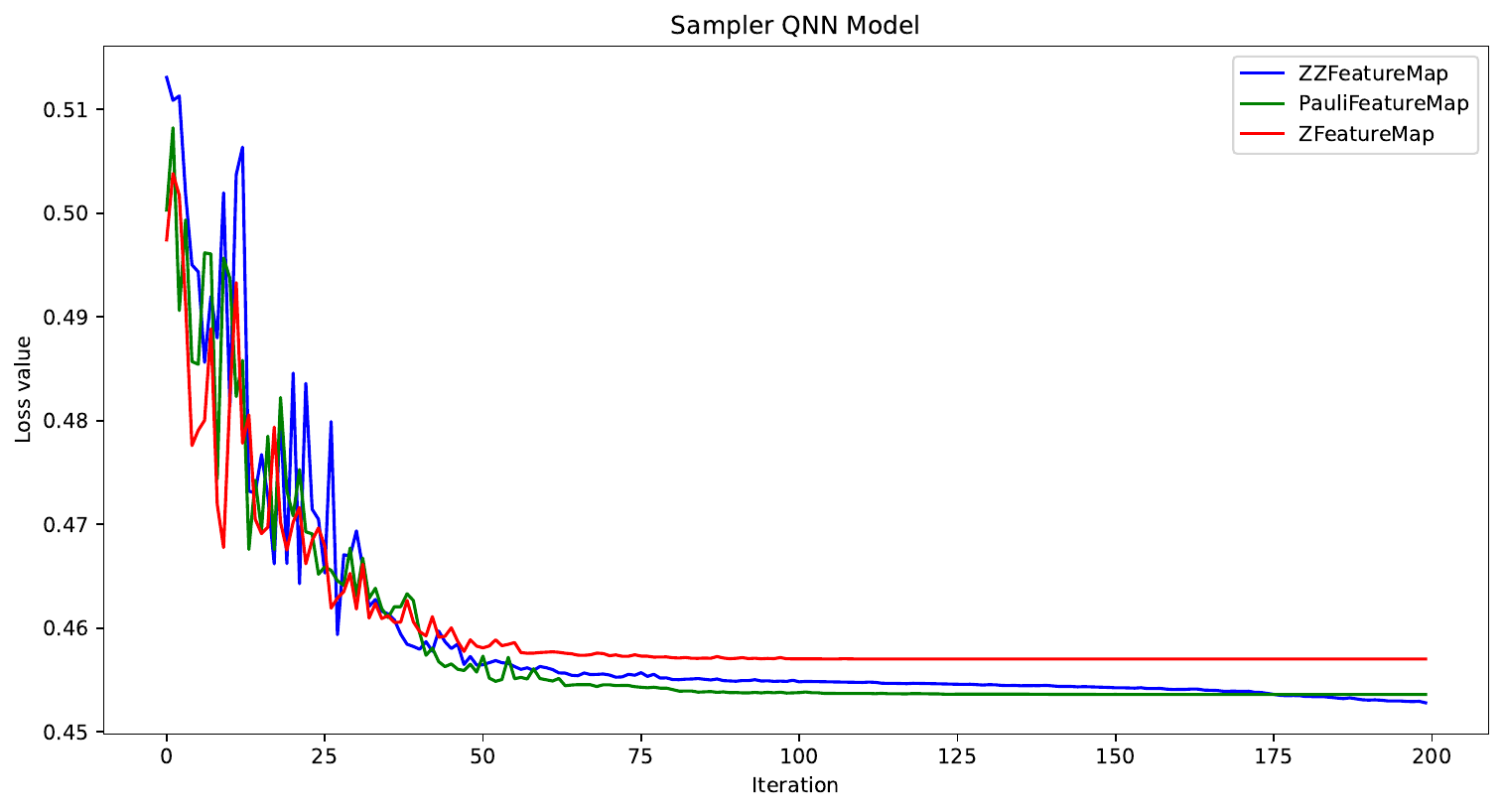} 
    \caption{Loss function of Sampler QNN model.}
    \label{fig:sqnnloss}
\end{figure}

\hyperlink{fig15}{Fig.~\ref{fig:sqnnloss}} shows the corresponding loss values, with the \texttt{ZFeatureMap} achieving a loss of $0.458$, the \texttt{PauliFeatureMap} a loss of 0$.454$, and the \texttt{ZZFeatureMap} a loss of $0.455$. The higher losses indicate that the SQNN model struggled to find an optimal solution, resulting in lower accuracy.
We also observed that the SQNN possesses lower accuracy than the other models, which aligns with our expectations because SQNNs are better suited to combinatorial optimisation and general constraint-imposing problems, such as scheduling problems, map colouring, and logic-placement number assignment games like Sudoku. The inherent limitations of SQNNs in handling continuous and high-dimensional data, as encountered in our dataset, could explain the observed lower accuracy in the context of fraud detection.

\section{Conclusion} \hypertarget{Concl}{}

In conclusion, our research presents a rigorous and insightful comparative study of four cutting-edge Quantum Machine Learning models: QSVC, VQC, EQNN, and SQNN. We have comprehensively understood their capabilities and limitations by evaluating their performance on a meticulously curated dataset and utilizing three distinct feature maps, \texttt{ZZFeatureMap}, \texttt{PauliFeatureMap}, and \texttt{ZFeatureMap}.

Among the models evaluated, QSVC stood out as the top performer, showcasing unparalleled excellence with F1 scores of $0.98$ for both fraud and non-fraud classes. Its utilisation of the Quantum kernel for state similarity measurement proves to be a potent strategy, circumventing the need for conventional loss functions and yielding extraordinary results.

VQC also demonstrated remarkable performance, boasting an impressive F1 score of $0.90$. However, we observed a potential area for refinement during its training process, suggesting avenues for future exploration to harness its power.
In contrast, EQNN and SQNN exhibited comparatively lower F1 scores, hinting at the influence of the Quantum circuits used for data encoding on their accuracy. Addressing these limitations might be the key to unlocking their potential in this field.

Our findings reinforce the promise of Quantum computing in revolutionizing machine learning paradigms. The exceptional performance of QSVC and VQC attests to the vast potential of Quantum algorithms for solving complex classification problems with unprecedented precision.
\newpage
\section*{Declarations}
\subsection*{Conflicts of interest}
The authors have no competing interests or other interests that might be perceived to influence the results and/or discussion reported in this paper.

\end{document}